\documentclass[12pt,a4paper]{article}
\usepackage{mathrsfs}
\usepackage{epsfig}

\usepackage{tabularx}
\usepackage{graphics}
\usepackage{graphicx}
\usepackage{float}
\usepackage{multirow}
\usepackage{tabularx}
\usepackage{epstopdf}
\usepackage{subfigure}

\begin{document}
\title{Constraints on the  charged-current non-standard neutrino interactions induced by the gauge boson $W'$}
\author{ Chong Xing Yue\thanks{E-mail:cxyue@lnnu.edu.cn} \  and  Xue Jia Cheng\thanks{E-mail:Cxj225588@163.com} \\
{\small Department of Physics, Liaoning  Normal University, Dalian
116029, P. R. China}}
\date{\today}

\maketitle
\begin{abstract}

 \vspace{0.5cm}
Many new physics scenarios predict the existence of the extra charged gauge boson $ W'$, which can induce the charged-current (CC) non-standard neutrino interactions (NSI).  We investigate the  constraints on the CC NSI in model-independent fashion via considering the $ W'$ contributions to the lepton flavor violating (LFV) decays $\ell_{i}\rightarrow \ell_{j}\gamma$, the pure leptonic flavor conservation (FC) decays $\ell_{i}\rightarrow \ell_{j}\nu_{i}\bar{\nu_{j}}$, leptonic decays of charged pion meson, semileptonic $\tau$ decays, and superallowed  $\beta$ decays. We find that the constraints on the pure leptonic CC NSI are generally stronger than the ones for the CC NSI with first generation quarks. The most stringent constraints on the CC NSI arise from the LFV decay $\mu \rightarrow e\gamma$. The constraints on the CC NSI from direct searches at the LHC are also discussed.

 \end{abstract}
\newpage

\noindent{\bf 1. Introduction }\vspace{0.5cm}

The standard model (SM) has achieved great success in describing the elementary particles as well as the fundamental interactions. However there are several well-known problems about phenomena the SM cannot explain. Among them, the existence of non-zero neutrino mass [1] is the most robust one, which has been firmly established by various experimental searches. The origin of neutrino mass clearly requires new physics beyond the SM (BSM) [2], which often comes with the new Fermi-type interactions of neutrinos with first generation fermions affecting neutrino production, propagation and detection processes, usually called as non-standard neutrino interactions (NSI) [3]. So far there are many theoretical and experimental efforts to understand all possible kinds of NSI effects and  search for new physics, for reviews see Refs.[4, 5].

NSI are new vector contact interactions between neutrinos and first generation fermions induced by BSM, which can be charged-current (CC) NSI inducing non-standard CC production and detection mechanisms for neutrinos or neutral-current (NC) NSI leading to new NC interactions with the SM fermions. NSI have made contributions to some observable quantities, such as particle decay rates, scattering cross-sections, neutrino mixing parameters, and matter-affected oscillation probabilities. The relevant experimental data might provide constraints on NSI [4, 5]. While CC NSI are more severely constrained [6, 7].

Many BSM scenarios predict the existence of the extra charged, massive, colorless gauge boson, usually called $ W'$ [1], which arises from general extensions of the SM gauge groups and might solve some phenomenological anomalies depending on its coupling strength and mass. So far, a large variety of works about the new gauge boson  $ W'$ have been performed, such as rare decays, muon anomalous magnetic moment, electroweak precision tests, and direct searches at the LHC. The constraints on its mass and gauge coupling are investigated.  The ATLAS and CMS collaborations  have set lower limits on the $ W'$ mass  $M_{W'}$ [8]. Searching  for this kind of new particles is and will continue to be one integral part of the present and future high energy collider physics programs. In this paper, we will focus our attention on the constraints on the CC NSI induced by the new gauge boson $ W'$.

To obtain the constraints on CC NSI, we will calculate the contributions of $W'$ to some low-energy observables in model-independent fashion and compare with the recent experimental values.  We mainly consider the lepton flavor violating (LFV) decays $\ell_{i}\rightarrow \ell_{j}\gamma$, the pure leptonic flavor conservation (FC) decays $\ell_{i}\rightarrow \ell_{j}\nu_{i}\bar{\nu_{j}}$, leptonic decays of charged pion meson, semileptonic $\tau$ decay, the lepton flavor universality (LFU) in pion meson and tau lepton decays, the CKM unitarity and superallowed  $\beta$ decays.

The rest of the paper is organized as follows. In section 2, we provide an overview of the theoretical framework and give the general form of the CC NSI strength induced by $ W'$.  The constraints on the pure leptonic CC NSI and the CC NSI with first generation quarks are investigated in sections 3 and 4, respectively. Our conclusions and discussions are given in section 5.

 \vspace{0.5cm} \noindent{\bf 2. New charged gauge boson $W'$ and CC NSI}

\vspace{0.5cm}As commented in introduction, NSI are new vector interactions between neutrinos and matter fields, which are induced by either a vector or charged-scalar mediator, divided into CC NSI and NC NSI, and can be parameterized in terms of the low-energy effective four-fermion Lagrangian. For the CC NSI interested in this paper, they can also be divided into two categories, the pure leptonic CC NSI and the CC NSI with first generation quarks [6,  7], which are respectively described by
\begin{eqnarray}
\mathcal{L}^{\ell}_{NSI}&=& -2{\sqrt{2}}{G_{F}}\varepsilon^{\ell\ell'x}_{\alpha\beta}[\bar{\ell}\gamma^{\mu}P_{x}\ell'][\bar{\nu_{\alpha}}\gamma_{\mu}P_{L}\nu_{\beta}],
\end{eqnarray}
and
\begin{eqnarray}
\mathcal{L}^{q}_{NSI}&=&
-2{\sqrt{2}}{G_{F}}\varepsilon^{udx}_{\alpha\beta}V_{ud}[\bar{u}\gamma^{\mu}P_{x}d][\bar{
\ell_{\alpha}}\gamma_{\mu}P_{L}\nu_{\beta}]+h.c..
\end{eqnarray}
Where  $G_F$ is the Fermi constant, $P_{x}$ with $x=L$ or $ R$ is either a left-handed or a right-handed projection operator, $\alpha$ and $\beta$ are lepton flavor indices. ${V}_{qq'}$ is the Cabbibo-Kobayashi-Maskawa (CKM) matrix element. We have assumed that there are only left-handed neutrinos. The parameters $\varepsilon^{ijx}_{\alpha\beta}$ are dimensionless coefficients that quantify the strength of the new vector interactions, which can be thought of in a simplified model framework as  $\varepsilon \sim g^{2}_{X} / M^{2}_{X} $ with $X$ being mediator. Due to Hermiticity, there is $\varepsilon^{ijx}_{\alpha\beta}=\varepsilon^{jix\ast}_{\beta\alpha}$. For the pure leptonic CC NSI, $ \ell\neq\ell'$ denotes a SM charged lepton.

CC NSI can affect neutrino production and detection at the relevant neutrino experiments and change the flavor distribution of the initial neutrino flux. The effects of CC NSI on reactor and long-baseline neutrino experiments have been discussed in literatures [for example see, Refs. [9, 10]]. Reference [11] has carefully analyzed the effects of CC NSI on neutrino oscillation experiments in framework of the effective field theory. References [6, 7] have investigated the bounds on the CC NSI parameters $\varepsilon^{ijx}_{\alpha\beta}$ in model-independent way. In this work, we will concentrate on the contributions of the new charged gauge boson $W'$ to some low-energy observables, compare with the recent experimental values and try to give the constraints on the CC NSI induced by $W'$ exchange.

New massive charged gauge boson $W'$ is predicted in many new physics scenarios. To ensure model independence, it is very convenient to describe the $W'$
 chiral couplings to the SM fermions by the general Lorentz invariant Lagrangian [12], which can be written as
\begin{eqnarray}
\mathcal{L}_{W'}&=&
-\frac{g}{\sqrt{2}}[V_{qq'}\bar{q}\gamma^{\mu}({A}^{qq'}_{L}P_{L}+{A}^{qq'}_{R}P_{R})q'+{B}^{\alpha\beta}_{L}\bar{\ell_{\alpha}}\gamma^{\mu}P_{L}\nu_{\beta}]{W}^{'}_{\mu}+h.c..
\end{eqnarray}
Where $g$ is the SM electroweak coupling constant, the parameters ${A}^{qq'}_{L,R}$ and ${B}^{\alpha\beta}_{L}$ are overall normalized by the weak coupling strength. For the SM gauge boson W, there should be ${A}^{qq'}_{L}={B}^{\alpha\beta}_{L}=1$,
${A}^{qq'}_{R}={0}$\renewcommand{\thefootnote}{\fnsymbol{footnote}} \footnote{Ref.[13] has studied the constraints on the right-handed couplings of the SM boson $W$ with fermions in the framework of EFT.}.

As long as there is no new source of flavor violation with respect to the SM, Eq.(3) can be used to describe any specific new physics model. The new massive boson $W'$ can mix with the SM boson $W$ in principle. Considering that the mixing angle $\xi$ is usually constrained to be small, $|\xi|\leq6\times10^{-4}$  [14], we will ignore this mixing effects in all discussions of this paper.

There are many works to investigate the possibility of discovery the new gauge boson $W'$ via direct and indirect searches. Lower bounds on its mass $M_{W'}$  are obtained, which almost depend on the fermionic couplings of the boson $W'$. So the mass and coupling constant of the new charged boson $W'$ are not independent free parameters. Its contributions to the CC NSI parameters $\varepsilon^{ijx}_{\alpha\beta}$ can be written as
\begin{eqnarray}
\varepsilon^{\ell\ell'L}_{\alpha\beta}=
{B}^{\ell\beta}_{L}{B}^{\ell'\alpha\ast}_{L}(\frac{M_{W}}{M_{W'}})^{2},\quad
\varepsilon^{\ell\ell'R}_{\alpha\beta}=0;
\end{eqnarray}
\begin{eqnarray}
\varepsilon^{udL}_{\alpha\beta}=
{A}^{ud}_{L}{B}^{\beta\alpha\ast}_{L}(\frac{M_{W}}{M_{W'}})^{2},\quad
\varepsilon^{udR}_{\alpha\beta}={A}^{ud}_{R}{B}^{\beta\alpha\ast}_{L}(\frac{M_{W}}{M_{W'}})^{2}.
\end{eqnarray}

 In following sections, we will consider the contributions of the new charged gauge boson $W'$ to some low-energy processes and give the constraints on the relevant parameters $\varepsilon^{ijx}_{\alpha\beta}$ from the recent experimental measured values.

\vspace{0.5cm} \noindent{\bf 3. Constraints on the pure  leptonic CC NSI}

\vspace{0.5cm} \noindent{\bf 3.1 The LFV process $\ell_{i}\rightarrow \ell_{j}\gamma$}

\vspace{0.5cm}Assuming conservation of charge and Lorentz invariance, the effective Lagrangian for the LFV process $\ell_{i}\rightarrow \ell_{j}\gamma$ can be written as
\begin{eqnarray}
\mathcal{L}&=&
\frac{e{A}^{M}_{ij}}{4}{m_{i}}\bar{\ell_{i}}\sigma^{\mu\nu}\ell_{j}{F_{\mu\nu}}+\frac{ie{A}^{E}_{ij}}{4}{m_{i}}\bar{\ell_{i}}\gamma^{\mu}\sigma^{\mu\nu}\ell_{j}{F_{\mu\nu}},
\end{eqnarray}
where we have assumed ${m_{i}}\gg{m_{j}}$ and neglected the contributions proportional to the lepton mass ${m_{j}}$. $F_{\mu\nu}$ denotes the field strength tensor of the electromagnetic field. ${A}^{M}_{ij}$ and ${A}^{E}_{ij}$ are related to the anomalous magnetic moment ${a_{\ell_{i}}}$ and electric dipole moment ${d_{\ell_{i}}}$, respectively. Then the branching ratio ${Br}(\ell_{i}\rightarrow \ell_{j}\gamma)$ can be general given by [15]
\begin{eqnarray}
{B r}(\ell_{i}\rightarrow \ell_{j}\gamma)&=&
\frac{3(4\pi)^{3}\alpha_{e}}{4G^{2}_{F}}(|{A}^{M}_{ji}|^{2}+|{A}^{E}_{ji}|^{2}){B r}(\ell_{i}\rightarrow \ell_{j}\nu_{i}\bar{\nu_{j}}).
\end{eqnarray}
${Br}(\ell_{i}\rightarrow \ell_{j}\nu_{i}\bar{\nu_{j}})$ denotes the branching ratio of the flavor conservation $($FC$)$ process $\ell_{i}\rightarrow \ell_{j}\nu_{i}\bar{\nu_{j}}$.

Using the coupling ${W'}\ell_{\alpha}\nu_{\beta}$ given by Eq.(3), one can obtain the expression form of the branching ratio $Br(\ell_{i}\rightarrow \ell_{j}\gamma)$ induced by the new charged gauge boson ${W'}$ at one loop
\begin{eqnarray}
{B r}(\ell_{i}\rightarrow \ell_{j}\gamma)&=&
\frac{25\alpha_{e}}{48\pi}(\frac{M_{W}}{M_{W'}})^{4}\sum\limits_{\alpha}|{B}^{i\alpha}_{L}{B}^{j\alpha\ast}_{L}|^{2}{B r}(\ell_{i}\rightarrow \ell_{j}\nu_{i}\bar{\nu_{j}})
\nonumber\\&=& \frac{25\alpha_{e}}{48\pi}\sum\limits_{\alpha}|\varepsilon^{ijL}_{\alpha\alpha}|^{2}{B r}(\ell_{i}\rightarrow \ell_{j}\nu_{i}\bar{\nu_{j}}).
\end{eqnarray}

Their current experimental upper limits are [1]:
\begin{eqnarray}
{B r}(\mu\rightarrow e \gamma)\leq 4.2\times10^{-13},\quad {B r}(\tau\rightarrow \mu\gamma)\leq 4.4\times10^{-8},
\end{eqnarray}
\begin{eqnarray}
{B r}(\tau\rightarrow e \gamma)\leq 3.3\times10^{-8}
\end{eqnarray}
at $90 \%$ confidence level (CL). Using above limits and the corresponding experimental measured values ${B r}(\mu\rightarrow e \nu_{\mu}\bar{\nu_{e}})\approx 1$, ${B r}(\tau\rightarrow \mu \nu_{\tau}\bar{\nu_{\mu}})=  (17.39\pm0.04)\%$, and ${B r}(\tau\rightarrow e \nu_{\tau}\bar{\nu_{e}})= (17.82\pm0.04)\%$ [1], we obtain the following bounds
\begin{eqnarray}
|\varepsilon^{\mu e L}_{\alpha\alpha}| \leq 1.076\times10^{-5},\quad
|\varepsilon^{\tau\mu L}_{\alpha\alpha}| \leq 8.359\times10^{-3},\quad
|\varepsilon^{\tau e L}_{\alpha\alpha}| \leq 7.151\times10^{-3}.
\end{eqnarray}

\vspace{0.5cm} \noindent{\bf 3.2 The pure leptonic FC decay $\ell_{i}\rightarrow \ell_{j}\nu_{i}\bar{\nu_{j}}$}

\vspace{0.5cm}It is well known that the pure leptonic FC decay $\ell_{i}\rightarrow \ell_{j}\nu_{i}\bar{\nu_{j}}$ proceeds through the exchange of the electroweak gauge boson $W$ in the SM. Including the contributions of the new charged gauge boson $W'$, the decay width can be given by
\begin{eqnarray}
\Gamma(\ell_{i}\rightarrow \ell_{j}\nu_{i}\bar{\nu_{j}})&=&
\Gamma^{SM}(\ell_{i}\rightarrow \ell_{j}\nu_{i}\bar{\nu_{j}})\Big \{|1+(\frac{M_{W}}{M_{W'}})^{2}{B^{ii}_{L}}{B^{jj\ast}_{L}}|^{2}
+(\frac{M_{W}}{M_{W'}})^{4}\sum\limits_{\alpha,\beta}|{B^{i \beta}_{L}}{B^{j \alpha\ast}_{L}}|^{2}\Big \}
\nonumber\\&=&\Gamma^{SM}(\ell_{i}\rightarrow \ell_{j}\nu_{i}\bar{\nu_{j}})[|1+\varepsilon^{ijL}_{ji}|^{2}+\sum\limits_{\alpha,\beta}|\varepsilon^{ijL}_{\alpha\beta}|^{2}]
\end{eqnarray}
with
\begin{eqnarray}
\Gamma^{SM}(\ell_{i}\rightarrow \ell_{j}\nu_{i}\bar{\nu_{j}})&=&
\frac{G^{2}_{F}m^{5}_{\ell}}{192\pi^{3}}(1+\Delta{q}).
\end{eqnarray}
Here $\Delta{q}$ includes phase space, QED and hadronic radiative corrections. The last term of Eq.(12) only comes from $W'$ exchange induced by the LFV coupling ${W'}\ell \nu_{\ell'}(\ell \neq \ell')$, which is the sum of  the flavor indexes $\beta $ and $\alpha $ in the case of $\beta \neq i$ and $\alpha \neq j $.

The Fermi constant $G_{F}$ has been determined with a very high precision from the muon lifetime [16]. The existence of the new charged gauge boson ${W'}$ would affect the determination of the $G_{F}$ value from muon decay.

Using Eq.(12) we have
\begin{eqnarray}
G^{2}_{F}&=&(G^{SM}_{F})^{2}[|1+\varepsilon^{\mu e L}_{e \mu}|^{2}+\sum\limits_{\alpha,\beta}|\varepsilon^{\mu e L}_{\alpha\beta}|^{2}],
\end{eqnarray}
where $G^{SM}_{F}$ does not contain any NP contributions and is expressed by the SM fundamental parameters
\begin{eqnarray}
G^{SM}_{F}&=&\frac{\pi\alpha_{e}M^{2}_{Z}}{\sqrt{2}M^{2}_{W}(M^{2}_{Z}-M^{2}_{W})(1-\Delta{r})}.
\end{eqnarray}
$ \Delta{r}$ includes the contributions from higher-order corrections (for example, see [17] and references therein), which depends on the relevant SM mass parameters.

\begin{table}[h]
\caption {The values and current uncertainties of the input parameters.}
\vspace{0.5cm}
\begin{center}
\begin{tabular}
{|p{2cm}|p{6cm}<{\centering}|}
 \hline
    Parameters & Experimental values  \\
\hline
    \(G_{F}\) & \(1.1663787(6)\times10^{-5}GeV^{-2}\)[1] \\
  \hline
    \(\alpha_{e}\) & \(7.2973525664(17)\times10^{-3}\)[1] \\
  \hline
    \(M_{Z}\) & \(91.1875\pm0.0021 GeV\)[1] \\
  \hline
    \(M_{W}\) & \(80.379\pm0.012 GeV [18]\) \\
  \hline
\end{tabular}
\end{center}
\end{table}

Using above equations, we obtain
\begin{eqnarray}
Re(\varepsilon^{\mu e L}_{e \mu}) =(-1.557\pm1.049)\times10^{-3},\quad
|\varepsilon^{\mu e L}_{\alpha\beta}| \leq0.032.
\end{eqnarray}
 The relevant input parameters used in our numerical calculation are given in Table 1. We have assumed only one non-zero $\varepsilon$ at a time in our numerical analysis. For any observable quantity, we have assumed that the experimental error and theoretical error are uncorrelated and have used the error propagation formula to obtain its uncertainty.

From Eq.(12) one can see that the experimental measured values of the branching ratios ${B r}(\ell_{i}\rightarrow \ell_{j}\nu_{i}\bar{\nu_{j}})$ can be used to constrain the $W'$ effects and further give bounds to the corresponding coupling parameters for the pure leptonic CC NSI. To do this, it is very convenient to define the relative correction parameter as
\begin{eqnarray}
R_{i}&=&\frac{\Gamma_{i}-\Gamma^{SM}_{i}}{\Gamma^{SM}_{i}}.
\end{eqnarray}
Where $\Gamma^{SM}_{i}$ denotes the SM prediction width and $\Gamma_{i}$ includes the contributions from both the SM and NP. The parameter $R_{i}$ is remarkably clean and very sensitive to new physics effects, as hadronic theory uncertainties are canceled. Combining Eq.(12) and Eq.(17), there are
\begin{eqnarray}
R(\tau\rightarrow \mu\nu_{\tau}\bar{\nu_{\mu}})&=&
2Re(\varepsilon^{\tau\mu L}_{\mu\tau})+\sum\limits_{\alpha,\beta}|\varepsilon^{\tau\mu L}_{\alpha\beta}|^{2},
\end{eqnarray}
\begin{eqnarray}
R(\tau\rightarrow e \nu_{\tau}\bar{\nu_{e}})&=&
2Re(\varepsilon^{\tau e
L}_{e \tau})+\sum\limits_{\alpha,\beta}|\varepsilon^{\tau e L}_{\alpha\beta}|^{2}.
\end{eqnarray}

\begin{table}[h]
\caption {The bounds on the NSI parameters $\varepsilon^{\tau jL}_{\alpha\beta}$ from the FC process $ \tau\rightarrow l_{j} \nu_{\tau}\bar{\nu_{j}} $.}
\vspace{0.5cm}
\begin{center}
\begin{tabular}
{|p{2cm}|p{6cm}<{\centering}|p{6cm}<{\centering}|}
 \hline
      &\(\tau\rightarrow \mu\nu_{\tau}\bar{\nu_{\mu}}\) & \(\tau\rightarrow e \nu_{\tau}\bar{\nu_{e}}\)  \\
  \hline
   \(Re(\varepsilon^{\tau\mu L}_{\mu\tau})\) & \((2.920\pm1.445)\times10^{-3}\) & \--\  \\
  \hline
   \(|\varepsilon^{\tau\mu L}_{\alpha\beta}|\) & \(0.055\) & \--\  \\
  \hline
   \(Re(\varepsilon^{\tau e L}_{e \tau})\) & \--\ & \((1.215\pm1.418)\times10^{-3}\)   \\
  \hline
   \(|\varepsilon^{\tau e L}_{\alpha\beta}|\) &\--\ & \(0.045\)  \\
  \hline
\end{tabular}
\end{center}
\end{table}

The decay width $\Gamma(\tau\rightarrow l  \nu_{\tau}\bar{\nu_{l}})$ has been calculated with high accuracy [19, 20]. Using the experimental measured values of the branching ratio
${Br}^{e x p}(\tau\rightarrow l \nu_{\tau}\bar{\nu_{l}})$ and the tau lifetime $\tau_{\tau}=(290.3\pm0.5)\times10^{-15}s$ [1], we obtain the bounds on the individual NSI parameters $\varepsilon^{\tau jL}_{\alpha\beta}$, which are shown in Table 2. From Table 2 one can see that these bounds are comparable to the bounds on $\varepsilon^{\mu e L}_{\alpha\beta}$ obtained from the kinematic determination of the Fermi constant.

The lepton flavor universality  (LFU ) in the pure leptonic decays of the charged leptons  is an excellent way to probe new physics.
Recently, the HFLAV collaboration reported three ratios from the pure leptonic decays $\ell_{i}\rightarrow \ell_{j}\nu_{i}\bar{\nu_{j}}$ [21]
\begin{eqnarray}
\frac{g_{\tau}}{g_{\mu}}=1.0010\pm 0.0014,\quad
\frac{g_{\tau}}{g_{e}}= 1.0029\pm 0.0014,
\end{eqnarray}
\begin{eqnarray}
\frac{g_{\mu}}{g_{e}}= 1.0018\pm 0.0014.
\end{eqnarray}

Neglecting the contributions of the LFV coupling ${W'}\ell \nu_{\ell'}$ with $\ell\neq\ell'$, we have
\begin{eqnarray}
\frac{g_{\tau}}{g_{\mu}}&=&\sqrt{\frac{\bar{\Gamma}(\tau\rightarrow e \nu_{\tau}\bar{\nu_{e}})}{\bar{\Gamma}(\mu\rightarrow e \nu_{\mu}\bar{\nu_{e}})}}
 =1+Re(\varepsilon^{\tau e L}_{e \tau}-\varepsilon^{\mu e L}_{e \mu}),
\end{eqnarray}
\begin{eqnarray}
\frac{g_{\tau}}{g_{e}}&=&\sqrt{\frac{\bar{\Gamma}(\tau\rightarrow \mu\nu_{\tau}\bar{\nu_{\mu}})}{\bar{\Gamma}(\mu\rightarrow e \nu_{\mu}\bar{\nu_{e}})}}
=1+Re(\varepsilon^{\tau\mu L}_{\mu\tau}-\varepsilon^{\mu e L}_{e \mu}),
\end{eqnarray}
\begin{eqnarray}
\frac{g_{\mu}}{g_{e}}&=&\sqrt{\frac{\bar{\Gamma}(\tau\rightarrow \mu\nu_{\tau}\bar{\nu_{\mu}})}{\bar{\Gamma}(\tau\rightarrow e \nu_{e}\bar{\nu_{\tau}})}}
=1+Re(\varepsilon^{\tau\mu L}_{\mu\tau}-\varepsilon^{\tau e L}_{e \tau}).
\end{eqnarray}
Here $\bar{\Gamma}$ denotes the decay width normalized to its SM prediction. Then we obtain the bounds on the deviations between different NSI parameters
\begin{eqnarray}
|Re(\varepsilon^{\tau e L}_{e \tau}-\varepsilon^{\mu e L}_{e \mu})|\leq 0.0024,\quad
|Re(\varepsilon^{\tau \mu L}_{\mu \tau}-\varepsilon^{\mu e L}_{e \mu})|\leq 0.0043,
\end{eqnarray}
\begin{eqnarray}
|Re(\varepsilon^{\tau \mu L}_{\mu \tau}-\varepsilon^{\tau e L}_{e \tau})|\leq 0.0032.
\end{eqnarray}

If  one assumes that the FC process $ \tau\rightarrow l \nu_{\tau}\bar{\nu_{l}} $ is not affected by $W'$ exchange, the bounds on the individual NSI parameter $\varepsilon^{\mu e L}_{\alpha\beta}$ can also be obtained from the
LFU ratio $g_{\tau}/g_{\mu}$ or  $g_{\tau}/g_{e}$, which are similar to those of Ref. [7].

\vspace{0.5cm} \noindent{\bf 4. Constraints on the CC NSI with first generation quarks}

\vspace{0.5cm} \noindent{\bf 4.1 The charged pion decay $\pi^{+}\rightarrow \ell^{+}{\nu}$}

\vspace{0.5cm}The decay process $P^{+}\rightarrow \ell^{+}{\nu}$ with $P$ being pseudoscalar meson is helicity suppressed in the SM, which is sensitive to new physics (for example see Ref. [22]) and thus is of great interest as a probe for new physics. In the SM the leading-order decay width can be written as
\begin{eqnarray}
\Gamma(P^{+}\rightarrow \ell^{+}_{\alpha}{\nu_{\alpha}})&=&
\frac{G^{2}_{F}}{8\pi}|V_{ij}|^{2}F_{P}^{2}m^{2}_{\ell}m_{P}(1-\frac{m^{2}_{\ell}}{m^{2}_{P}})^{2},
\end{eqnarray}
where $F_{P}$ and $m_{P}$ are the decay constant and mass of the charged pseudoscalar meson $P$, respectively.

Including the $W'$ contributions, the partial decay width $\Gamma(\pi^{+}\rightarrow \ell^{+}_{\alpha}{\nu_{\alpha}})$ is given by
\begin{eqnarray}
\Gamma(\pi^{+}\rightarrow \ell^{+}_{\alpha}{\nu_{\alpha}})&=&
\Gamma^{SM}(\pi^{+}\rightarrow \ell^{+}_{\alpha}\nu_{\alpha})\Big \{|1+(\frac{M_{W}}{M_{W'}})^{2}(A^{ud}_{R}-A^{ud}_{L})B^{\alpha\alpha\ast}_{L}|^{2}
\nonumber\\&+& (\frac{M_{W}}{M_{W'}})^{4}\sum\limits_{\beta}|(A^{ud}_{R}-A^{ud}_{L})B^{\alpha\beta\ast}_{L}|^{2}\Big \}
\nonumber\\&=& \Gamma^{SM}(\pi^{+}\rightarrow \ell^{+}_{\alpha}\nu_{\alpha})[|1+\varepsilon^{udA}_{\alpha\alpha}|^{2}+\sum\limits_{\beta}|\varepsilon^{udA}_{\alpha\beta}|^{2}].
\end{eqnarray}
Where the lepton flavor indices $\alpha\neq\beta=e, \mu$ and $\tau$. The SM prediction $\Gamma^{SM}(\pi^{+}\rightarrow \ell^{+}_{\alpha}{\nu_{\alpha}})$ contains the SM loop corrections [23], which is given by
\begin{eqnarray}
\Gamma(\pi^{+}\rightarrow \ell^{+}_{\alpha}{\nu_{\alpha}})&=&
\frac{G^{2}_{F}}{8\pi}|V_{ud}|^{2}F_{\pi}^{2}m^{2}_{\ell}m_{\pi}(1-\frac{m^{2}_{\ell}}{m^{2}_{\pi}})^{2}(1+\delta_{\pi}),
\end{eqnarray}
where $\delta_{\pi}$ parameterizes radiative corrections.

The leptonic decay of the charged pion meson, $\pi^{+}\rightarrow \ell^{+}_{\alpha}{\nu_{\alpha}}$, can be used as one of the most sensitive probes of the electron-muon flavor universality\renewcommand{\thefootnote}{\fnsymbol{footnote}}\footnote{Ref.[24] recently studied the constraints on the quark-lepton charged currents in general neutrino interactions with sterile neutrinos in the framework of EFT.}. The  ratio $R^{\pi}_{{e}/{\mu}}=\Gamma(\pi^{+}\rightarrow e^{+}{\nu_{e}})$ / $\Gamma(\pi^{+}\rightarrow \mu^{+}{\nu_{\mu}})$ has been calculated and measured with high accuracy. Compare the SM prediction $R^{\pi}_{{e}/{\mu}}=(1.2352\pm0.0001)\times10^{-4}$ [25] with the experimental world average $R^{\pi}_{{e}/{\mu}}=(1.2327\pm0.0023)\times10^{-4}$ [1], one can obtain the constraints on the corresponding parameters $\varepsilon^{udA}_{ij}$ of the CC NSI with the first generation  quarks via the following formula
\begin{eqnarray}
R^{\pi}_{{e}/{\mu}}&=&
(R^{\pi}_{{e}/{\mu}})^{SM}[1+2Re(\varepsilon^{udA}_{ee}-\varepsilon^{udA}_{\mu\mu})].
\end{eqnarray}
In above equation, the flavor non-diagonal contributions of the new gauge boson $W'$ have also neglected.
Then we have
\begin{eqnarray}
|Re(\varepsilon^{udA}_{ee}-\varepsilon^{udA}_{\mu\mu})| \leq 1.944\times10^{-3}.
\end{eqnarray}

We can only obtain the constraint on  the deviation $Re(\varepsilon^{udA}_{ee}-\varepsilon^{udA}_{\mu\mu})$ via the electron-muon flavor universality. However, the relative correction parameter $R_{i}=({\Gamma^{i}-\Gamma^{SM}_{i}})/{\Gamma^{SM}_{i}}$ can give the constraints on the individual NSI parameter $\varepsilon^{udA}_{ij}$ via the following relations
\begin{eqnarray}
R(\pi^{+}\rightarrow e^{+}{\nu_{e}})&=& 2Re(\varepsilon^{udA}_{ee})+\sum\limits_{\beta}|\varepsilon^{udA}_{e \beta}|^{2},
\end{eqnarray}
\begin{eqnarray}
R(\pi^{+}\rightarrow\mu^{+}{\nu_{\mu}})&=&2Re(\varepsilon^{udA}_{\mu\mu})+\sum\limits_{\beta}|\varepsilon^{udA}_{\mu \beta}|^{2}.
\end{eqnarray}
Our results are shown in Table 3. In our numerical analysis we have taken the $\pi^{\pm}$ lifetime $\tau_{\pi}=(26.033\pm 0.005)\times 10^{-9}s$, the branching ratios ${B r}^{exp}(\pi^{+}\rightarrow e^{+}{\nu_{e}})=(1.230\pm0.004)\times10^{-4}$ and ${B r}^{exp}(\pi^{+}\rightarrow \mu^{+}{\nu_{\mu}})=(99.98770\pm0.00004)\% $ [1]. Due to the larger deviation between the SM prediction and experimental measured values for the decay width $\Gamma ( \pi^{+}\rightarrow e^{+}{\nu_{e}})$, the weaker constraints on the NSI parameter $ \varepsilon^{udA}_{ee}$ are given.

\begin{table}[h]
\caption {The constraints on the NSI parameters $\varepsilon^{u d A}_{\alpha\beta}$ from the decay $\pi^{+}\rightarrow \ell^{+}{\nu_{\ell}}$.}
\vspace{0.5cm}
\begin{center}
\begin{tabular}
{|p{2cm}|p{4cm}<{\centering}|p{4cm}<{\centering}|}
 \hline
      &\(\pi^{+}\rightarrow e^{+}{\nu_{e}}\) & \(\pi^{+}\rightarrow\mu^{+}{\nu_{\mu}}\)  \\
  \hline
   \(Re(\varepsilon^{udA}_{ee})\) & \(-0.0185\pm0.0090\) & \--\  \\
  \hline
   \(|\varepsilon^{udA}_{e\beta}|\) & \(0.138\) & \--\  \\
  \hline
   \(Re(\varepsilon^{udA}_{\mu \mu}) \) & \--\ & \(0.0025\pm0.0095\)   \\
  \hline
   \(|\varepsilon^{udA}_{\mu\beta}|\) &\--\ & \(0.118\)  \\
  \hline
\end{tabular}
\end{center}
\end{table}

 From above discussions we can see that the experimental uncertainties for the decay widths $\Gamma(\pi^{+} \rightarrow \mu^{+}\nu_{\mu})$ and $\Gamma(\pi^{+} \rightarrow e^{+}\nu_{e})$ are very small, while their theoretical uncertainties mainly from the decay constant and radiative corrections are relatively large.  Reference [26] has shown the universal theoretical uncertainties as
 \begin{eqnarray}
 \frac{\delta\Gamma^{SM}(\pi^{+} \rightarrow e^{+}\nu_{e})}{\Gamma^{SM}(\pi^{+} \rightarrow e^{+}\nu_{e})}= \frac{\delta\Gamma^{SM}(\pi^{+} \rightarrow \mu^{+}\nu_{\mu})}{\Gamma^{SM}(\pi^{+} \rightarrow \mu^{+}\nu_{\mu})}=1.9 \times 10^{-2}.
 \end{eqnarray}
 If we neglect the experimental uncertainties and assume that the contributions of the new charged gauge boson $W'$ to the decay width of the decay process $\pi^{+}\rightarrow \ell^{+}_{\alpha}{\nu_{\alpha}}$ do not exceed the theoretical uncertainties, then we can obtain the constraints on the NSI parameters
\begin{eqnarray}
Re(\varepsilon^{udA}_{ee})= Re(\varepsilon^{udA}_{\mu\mu}) \leq 9.5\times10^{-3}, \quad
|\varepsilon^{udA}_{e\beta}|=|\varepsilon^{udA}_{\mu\beta}|\leq 0.0796.
\end{eqnarray}

\vspace{0.5cm} \noindent{\bf 4.2 Semileptonic $\tau$ decays}

\vspace{0.5cm}The lepton $\tau$ is the only observed lepton heavy enough to decay into hadrons, which can be used to learn about fundamental physics [19]. In the SM, semileptonic $\tau$ decays are induced by exchange of the charged electroweak gauge boson $W$ connecting neutrino $\nu_{\tau}$ and quarks. These decays provide an ideal tool for testing the SM and NP effects in very clean conditions. The new charged gauge boson $W'$ has contributions to all of semileptonic $\tau$ decays in principle. The main purpose of this subsection is to investigate the constraints of semileptonic $\tau$ decays on the CC NSI with up and down quarks. So we focus our attention on the decay processes $\tau^{-}\rightarrow \pi^{-}\nu_{\tau}$, $\pi^{-}\pi^{0}\nu_{\tau}$ ,$\pi^{-}\eta^{(\prime)}\nu_{\tau}$ and $K^{-}K^{0}\nu_{\tau}$.

Including the contributions of the new charged gauge boson $W^{'}$, the expression forms  for the partial decay widths of above decay processes can be written as
\begin{eqnarray}
\Gamma^{SM+W'}(\tau^{-}\rightarrow \pi^{-}\nu_{\tau})&=&
\Gamma^{SM}(\tau^{-}\rightarrow \pi^{-}\nu_{\tau})
\Big \{|1+(\frac{M_{W}}{M_{W'}})^{2}(A^{ud}_{R}-A^{ud}_{L})B^{\tau\tau\ast}_{L}|^{2}
\nonumber\\&+& (\frac{M_{W}}{M_{W'}})^{4}\sum\limits_{\beta}|(A^{ud}_{R}-A^{ud}_{L})B^{\tau\beta\ast}_{L}|^{2}\Big \}
\nonumber\\&=& \Gamma^{SM}(\tau^{-}\rightarrow \pi^{-}\nu_{\tau})[|1+\varepsilon^{udA}_{\tau\tau}|^{2}+\sum\limits_{\beta}|\varepsilon^{udA}_{\tau\beta}|^{2}];
\end{eqnarray}
\begin{eqnarray}
\Gamma^{SM+W'}(\tau^{-}\rightarrow \pi^{-}\pi^{0}\nu_{\tau})&=&\Gamma^{SM}(\tau^{-}\rightarrow \pi^{-}\pi^{0}\nu_{\tau})
[|1+\varepsilon^{udV}_{\tau\tau}|^{2} +\sum\limits_{\beta}|\varepsilon^{udV}_{\tau\beta}|^{2}].
\end{eqnarray}
For the decays $\tau^{-}\rightarrow \pi^{-}\eta^{(\prime)}\nu_{\tau}$ and $\tau^{-}\rightarrow K^{-}K^{0}\nu_{\tau}$, the decay widths have similar forms with Eq.(37), only replace $\Gamma^{SM}(\tau^{-}\rightarrow \pi^{-}\pi^{0}\nu_{\tau})$ by the corresponding SM predictions. For the decay $\tau^{-}\rightarrow \pi^{-}\nu_{\tau}$, its SM prediction is described by the following simple form
\begin{eqnarray}
\Gamma^{SM}(\tau^{-}\rightarrow \pi^{-}\nu_{\tau})&=&
\frac{G^{2}_{F}|V_{ud}|^{2}F^{2}_{\pi}m^{3}_{\tau}}{16\pi}(1-\frac{m^{2}_{\pi}}{m^{2}_{\tau}})^{2}(1+\delta).
\end{eqnarray}
Where $\delta$ accounts from higher-order corrections [27].

The decay channel $\tau^{-}\rightarrow \pi^{-}\nu_{\tau}$ can also be used to test university via the parameter
$R_{{\tau}/{\pi}}={\Gamma(\tau^{-}\rightarrow \pi^{-}\nu_{\tau})}/{\Gamma(\pi^{-}\rightarrow \mu^{-}\nu_{\mu})}$. Using the measured value $(g_{\tau}/g_{\mu})_{\pi}=0.9958\pm0.0026$ [21], we obtain
\begin{eqnarray}
|Re(\varepsilon^{udA}_{\tau\tau}-\varepsilon^{udA}_{\mu\mu})| \leq 0.0068,
\end{eqnarray}
which is slightly weaker than the constraint on  the deviation $Re(\varepsilon^{udA}_{ee}-\varepsilon^{udA}_{\mu\mu})$ via the electron-muon flavor universality.

References [28, 29, 30] have analyzed the sensitivity of some semileptonic $\tau$ decays to new physics effects in a model independent way and discussed the values of the correction parameter $R_{i}= (\Gamma_{i}-\Gamma^{SM}_{i})/\Gamma^{SM}_{i}$ for different decay channels. From their results we have to say that the parameter $R_{i}$ has larger relative uncertainty. All the same, these decay processes can also generate   constraints on the CC NSI parameters. For example, $R(\tau^{-}\rightarrow\pi^{-}\nu_{\tau})=(0.12\pm 0.68)\times10^{-2}$  [28] demands
\begin{eqnarray}
Re(\varepsilon^{udA}_{\tau\tau})=(0.600\pm3.398)\times10^{-3},\quad
|\varepsilon^{udA}_{\tau\beta}|\leq 0.094;
\end{eqnarray}
while $R(\tau^{-}\rightarrow \pi^{-}\pi^{0}\nu_{\tau})=(0.89\pm0.44)\times10^{-2}$ [29] gives
\begin{eqnarray}
Re(\varepsilon^{udV}_{\tau\tau})=(4.440\pm2.190)\times10^{-3},\quad
|\varepsilon^{udV}_{\tau\beta}|\leq 0.083.
\end{eqnarray}
We expect that, with the further development of experiment and theory, semileptonic $\tau$ decays would give more stringent constraints on  the CC NSI with up and down quarks.

\vspace{0.5cm} \noindent{\bf 4.3 Superallowed $\beta$ decays}

\vspace{0.5cm}Theoretical and experimental advances in recent years have made nuclear/neutron $\beta$ decays be important for high-precision test of the SM and sensitive to new physics effects [31]. At leading order, $\beta$ decays only involve the first generation fermions and proceed via the exchange of the charged electroweak gauge boson $W$ in the SM. Thus, BSM can contribute to these decays via additional couplings $Wff'$ or new contact four-fermion interactions.

Superallowed $\beta$ decays, long-lived nuclear $O^{+}\rightarrow O^{+}$ transitions, are pure Fermi transitions, and depend uniquely on the vector coupling of the weak interaction. Presently, the most precise determination of the CKM matrix element $V_{ud}$ is obtained from $O^{+}\rightarrow O^{+}$ decays for nuclei ranging from $^{10}C$ to $^{74}R_{b}$ [32]. Considering the contributions of the new gauge boson $W^{\prime}$ to these decays, we have
\begin{eqnarray}
|V^{\beta}_{ud}|^{2}&=&|V^{SM}_{ud}|^{2}\Big \{|1+(\frac{M_{W}}{M_{W'}})^{2}(A^{ud}_{R}+A^{ud}_{L})B^{ee\ast}_{L}|^{2}
+(\frac{M_{W}}{M_{W'}})^{4}\sum\limits_{\alpha}|(A^{ud}_{R}
\nonumber\\&+&  A^{ud}_{L})B^{e\alpha\ast}_{L}|^{2}\Big \}
\nonumber\\&=& |V^{SM}_{ud}|^{2}[|1+\varepsilon^{udV}_{ee}|^{2}+\sum\limits_{\alpha}|\varepsilon^{udV}_{e\alpha}|^{2}].
\end{eqnarray}
Where $V^{SM}_{ud}$ is determined by the SM unitary condition $|V_{ud}|^{2}+|V_{us}|^{2}+|V_{ub}|^{2}=1$. Since we only discuss the constraints of superallowed  $\beta$ decays to the CC NSI with the first generation quarks in this subsection, we have neglected the corrections of $W^{'}$ to the Fermi constant $G_{F}$ from pure leptonic interaction $W^{'}\ell\nu$ in above equation.

From above equation we can obtain
\begin{eqnarray}
2Re(\varepsilon^{udV}_{ee})+\sum\limits_{\alpha}|\varepsilon^{udV}_{e\alpha}|^{2}&=&
\frac{|V^{\beta}_{ud}|^{2}}{1-|V_{ub}|^{2}-|V_{us}|^{2}}-1.
\end{eqnarray}
The CKM matrix element $V_{us}$ can be precision determined from Kaon and tau  decays, which is not affected by the CC NSI with the first generation quarks. Its value can also be extracted from  superallowed  $\beta$ decays, which has been studied recently in Refs. [33, 34]. We will take its value as the average value from Kaon and tau  decays  $|V_{us}|=0.2240 \pm 0.0005$ [34]. Although the $V_{ub}$ value  is very small and can not generate a large affect on the determination of the CC NSI parameters $\varepsilon^{udV}_{e j}(j=e, \mu, \tau)$, we will use $|V_{ub}| \approx 0.003683$ [35].

The $V^{\beta}_{ud}$ value extracted from  superallowed  $\beta$ decays suffers from theoretical uncertainty arose from radiative corrections [36]. In our numerical analysis, we quote three recent results  $|V^{\beta}_{ud}|_{SGPR}=0.97370 \pm 0.00014$ [37], $|V^{\beta}_{ud}|_{CMS}=0.97389 \pm 0.00018$ [38], and $|V^{\beta}_{ud}|_{SFGJ}=0.97365 \pm 0.00015$ [39], which are consistent with each other and  there are
 \begin{eqnarray}
 \frac{|V^{\beta}_{ud}|_{SGPR}^{2}}{1-|V_{ub}|^{2}-|V_{us}|^{2}} = 0.99819\pm0.00037,\quad \frac{|V^{\beta}_{ud}|_{CMS}^{2}}{1-|V_{ub}|^{2}-|V_{us}|^{2}} = 0.99858\pm0.00044,
 \end{eqnarray}
 \begin{eqnarray}
 \frac{|V^{\beta}_{ud}|_{SFGJ}^{2}}{1-|V_{ub}|^{2}-|V_{us}|^{2}}= 0.99809\pm0.00039.
\end{eqnarray}

 Using these input values of the CKM elements, one can easily give the individual constraints on $\varepsilon^{udV}_{ee}$ and $\varepsilon^{udV}_{e \alpha}$ $(\alpha\neq e)$ from superallowed $\beta$ decays, which are shown in Table 4 and are consistent with the results given by Ref. [7] at $1\sigma $ range. Recently, Ref. [9] has studied the constraints on the CC NSI from beta-minus decay data, and their results are $|Re(\varepsilon^{udV}_{e e})|\leq 0.001$ and $ |\varepsilon^{udV}_{e \alpha}|\leq 0.04$ in the case of assuming the uncertainty in beta-minus decay width as $\delta \Gamma/\Gamma=0.001$, which are comparable to our results.

\begin{table}[h]
\caption {The constraints on the NSI parameters $\varepsilon^{u d V}_{\alpha\beta}$ from superallowed $\beta$ decays.}
\vspace{0.5cm}
\begin{center}
\begin{tabular}
{|p{1.5cm}|p{4cm}<{\centering}|p{4cm}<{\centering}|p{4cm}<{\centering}|}
 \hline
      &\(|V^{\beta}_{ud}|_{SGPR}\) & \(|V^{\beta}_{ud}|_{CMS}\) & \(|V^{\beta}_{ud}|_{SFGJ}\)  \\
  \hline
   \(Re(\varepsilon^{udV}_{ee})\) & \((-0.905\pm0.186)\times10^{-3}\) &\((-0.710\pm0.219)\times10^{-3}\) &\((-0.956\pm0.194)\times10^{-3}\)   \\
  \hline
   \(|\varepsilon^{udV}_{e \alpha}|\)& \(0.033\) & \(0.031\) & \(0.034\) \\
  \hline
\end{tabular}
\end{center}
\end{table}

\vspace{0.5cm} \noindent{\bf 5. Conclusions and discussions}

We first investigate the contributions of the extra charged gauge boson $W'$ to the LFV decays $\ell_{i}\rightarrow \ell_{j}\gamma$, the pure leptonic FC decays $\ell_{i}\rightarrow \ell_{j}\nu_{i}\bar{\nu_{j}}$, leptonic decays of charged pion meson, semileptonic $\tau$ decays,  and superallowed  $\beta$ decays in model-independent way. Then, considering the lepton flavor universality in charged pion meson and tau lepton decays, and the CKM unitarity, we  discuss the constraints on the CC NSI induced by the new gauge boson $ W'$ from low-energy precision measurements. Based on the corresponding experimental values, the bounds on the individual NSI parameter $\varepsilon^{ijx}_{\alpha\beta}$ and the deviations between different NSI parameters are obtained, which are around $\mathcal{O}(10^{-4})$ to $\mathcal{O}(10^{-1})$. We find that the constraints on the pure leptonic CC NSI are generally stronger than the ones for the CC NSI with first generation quarks. The most stringent constraints on the CC NSI arise from the LFV decay $\mu \rightarrow e\gamma$ at one loop.

It is well known that, since the discovery of the SM gauge boson W, searching for the extra charged gauge boson $W'$ has continued in high energy collider experiments. The constraints on its mass and gauge coupling are obtained, which can be translated into the bounds on the CC NSI parameters $\varepsilon^{ijx}_{\alpha\beta}$ from discussions given in section 2. The ATLAS and CMS collaborations have set lower limits on the $ W'$ mass  $M_{W'}$ in the case of assuming its couplings to fermions with the same strength as those of the SM boson W [8]. Then, using Eq.(5), one can estimate the constraints on the relevant NSI parameters from the results of Ref.[8]. The most stringent limits to date from the ATLAS data are $|\varepsilon^{u d L}_{e e}| \leq 1.8 \times10^{-4}$ and $|\varepsilon^{u d L}_{\mu \mu}| \leq 2.5 \times10^{-4}$. Of course, we only considered the contributions of the sub-process $u d \rightarrow W' \rightarrow \ell\nu$ to the process $p p \rightarrow W' \rightarrow \ell\nu$. If we also consider the contributions from other sub-processes, such as $c s \rightarrow W' \rightarrow \ell\nu$, then these restrictions will be relaxed. Therefore, the constraints on the relevant CC NSI parameters from the LHC data are comparable to the results from low-energy experimental data. Certainly, the precise limitations of the LHC data on the CC NSI parameters should be carefully studied and furthermore the ability of the future collider experiments, such as HL-LHC, to detect these parameters is needed to be investigated. We will carry out these in the near future.

\section*{ACKNOWLEDGMENT}
This work was partially supported by the National Natural Science Foundation of China under Grants No. 11875157 and Grant No. 11847303. X. J. Cheng would like to thank Chun Hua Li for very useful discussions.

\end{document}